\documentclass[twocolumn,showpacs,aps,amsmath,amssymb,preprintnumbers]{revtex4}

\newcommand{\BE}{\begin{equation}}
\newcommand{\EE}{\end{equation}}
\newcommand{\BA}{\begin{eqnarray}}
\newcommand{\EA}{\end{eqnarray}}
\usepackage{graphicx}
\usepackage{dcolumn}
\usepackage{bm}

\preprint{cond-mat/0211622}

\begin{document}

\title{Questions of perfect lenses by left handed materials}

\author{Zhen Ye}
\affiliation{Wave Phenomena Laboratory, Department of Physics,
National Central University, Chungli, Taiwan}

\date{\today}

\begin{abstract}

We consider questions about the much discussed ``perfect lenses"
made by left handed materials. The transmission and reflection
from a slab of left handed materials are investigated and the
coefficients are obtained by the standard transfer matrix method.
Possible limitations on such superlenses are explored. It is shown
that the quality of the lenses can be significantly affected by
the absorption that is necessarily present in the materials.

\end{abstract}

 \maketitle

The resolution of a traditional optical lens has been restricted
to the order of the wavelength of the illuminating wave. In the
year of 2000, this traditional limitation was challenged by
Pendry\cite{Pendry}. He proposed that a class of ``superlenses"
could be made by the so called Left Handed Materials (LHM) or
Negative Refraction Index Materials (NRIM), the concept first
introduced by Veselago many years ago\cite{Ves}. Such lenses may
overcome the traditional limitation and makes `perfect' images.
Since then, the research on such a super lens and LHM has been
booming. A great body of literature has been generated (e.~g. the
references cited in \cite{Valanju}).

Recently, however, the super lens phenomenon was questioned by a
number of authors\cite{Valanju,Hooft,comm,Garcia}. In
\cite{Valanju}, the authors suggested that there is something
fundamental wrong with the research on LHM. By a numerical
simulation, they showed that the experimental results\cite{exp} on
LHM were an artifact. The authors in \cite{Garcia}, on the other
hand, showed that although there is amplification of evanescent
waves in ideal lossless and dispersiveless media, this is limited
to a finite width of the lens so that a perfect restoration of the
illuminated object is impossible. In addition, they showed that
the necessary presence of absorption may change drastically the
amplification feature. Since the dispute about the perfect lenses
still goes on, it is desirable to explore the issue further and to
quantify the capability of the LHM made lenses in making images.

In this Letter, we re-examine the issue of super lenses.
Electromagnetic wave propagation through a slab of LHM will be
studied by the standard transfer matrix. To avoid any possible
ambiguity such as those with regard to the multiple scattering
theory in \cite{Pendry} raised up by 't Hooft\cite{Hooft}, we will
give a detailed derivation. As pointed out in \cite{Garcia}, it
can be shown that the ideal lossless LHM is not physical. We show
that the LHM lenses can indeed amplify the evanescent waves which
have been lost in the imaging by traditional lenses. However, the
necessary presence of absorption, even a small amount, plays a
crucial rule in recovery of the lost evanescent waves, thereby
controlling the quality of imaging, and makes LHM lenses less
perfect. In other words, no perfect lenses are possible. Some
ambiguities surrounding the issue are also discussed.

The wave field radiated from an object in vacuum can be described
by a Fourier expansion \BE \phi({\bf r}) = \int_{-\infty}^\infty
d{\bf k}_\perp A(k_\perp,k_z) e^{i{\bf k}_\perp\cdot{\bf r}_\perp
+ ik_zz},\label{eq:int}\EE with $$k_\perp^2 + k_z^2 =
\frac{\omega^2}{c_0^2},$$ with $c_0$ being the speed of light in
the vacuum. Clearly, the components in the integration for
$|k_\perp|
> \frac{\omega}{c_0}$ are evanescent along the propagating path
$z$. They will be lost along the path and therefore will not be
able to contribute to imaging. This restricts the maximum
resolution in imaging to $2\pi/(\omega/c_0) = \lambda$. To
overcome the problem, it was suggested \cite{Pendry} that by
propagating through a slab of LHM, the evanescent waves can be
amplified to the extend that they can be effectively recovered at
the imaging site, and therefore the slab of LHM reveals a ``super
lens". In this paper, we will examine this issue. The only
requirements in the discussion are: (1) the imaginary parts of the
permittivity and permeability cannot be zero simultaneously; (2)
the imaginary part of any wavenumber in any medium should be
positive. These two conditions are necessary for the causality
principle to be held.

\input{epsf}
\begin{figure}[hbt]
\begin{center}
\epsfxsize=2in \epsffile{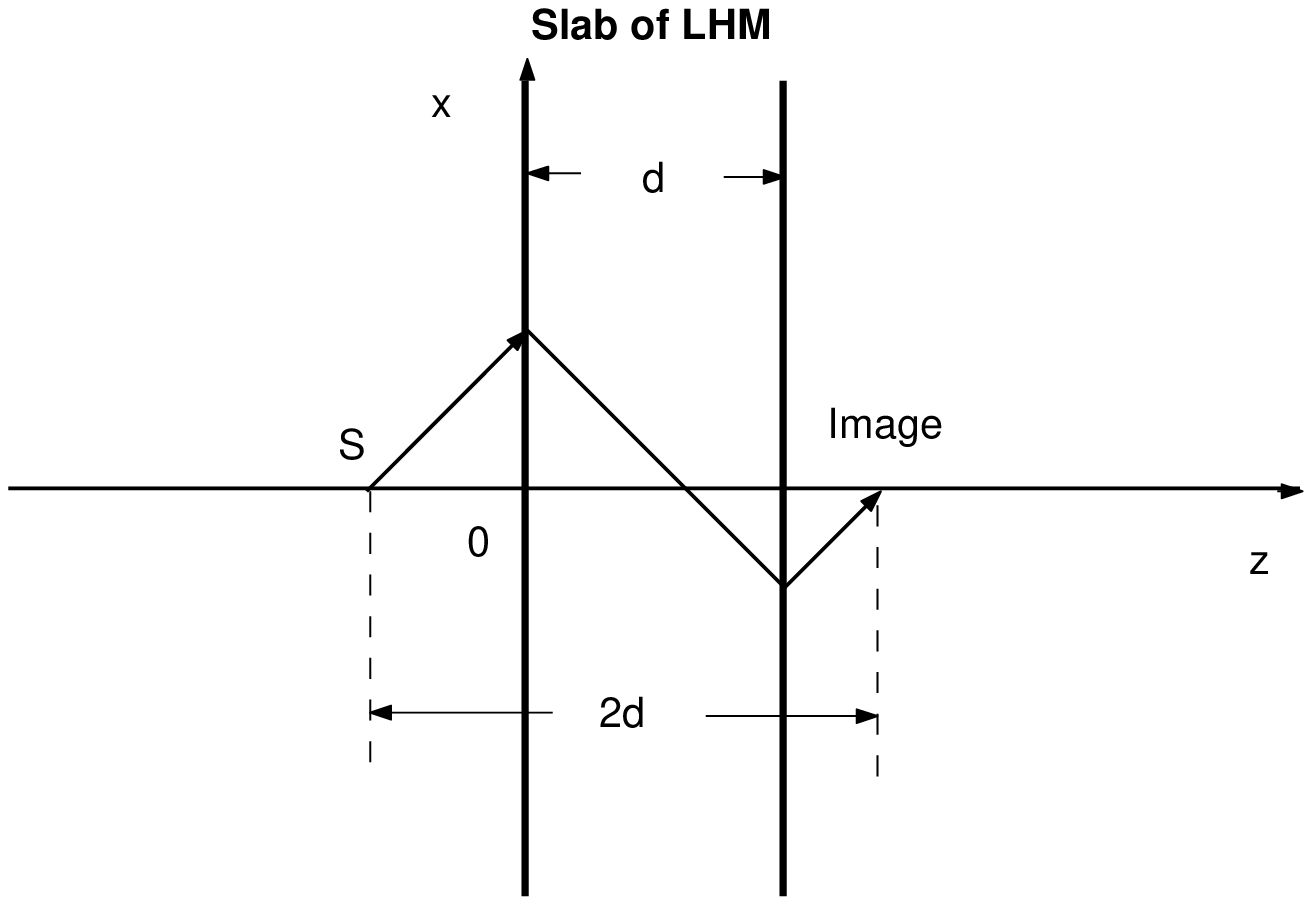} \caption{ \label{fig1}\small
The conceptual layout of the imaging from a slab of LHM.}
\end{center}
\end{figure}

Fig.~\ref{fig1} illustrates the concept of the imaging of a lens
made by LHM. Suppose that a slab of LHM is placed in the vacuum
and is located between $z=0$ and $z=d$; so the width of the slab
is $d$. The normal vector is along the $z$ axis. We define three
regions according to the $z$ values. Region I is for $z<0$, region
II covers $0<z<d$, i.~e. the region of the LHM, and region III is
$z>d$. An object denoted by `S' is located at some distance to the
left of the slab; the distance between the object and the lens
should be smaller than $d$. After the waves radiated from the
object pass the slab, they will make an image at distance $2d$
from the object.

For simplicity yet without losing generality, we consider a plane
wave incident from $z<0$ to the slab of LHM. Let the incidence
plane be the $x-z$ plane.

First, we determine the wavenumbers inside the slab. The wave
equation inside the slab is $$ \left(\nabla^2 - (\epsilon \mu)
\frac{\partial^2}{\partial t^2}\right) \left\{ \begin{array}{c}
E\\ H
\end{array}\right\} = 0,$$ leads to the dispersion\BE {k'}^2 + {k_x'}^2 =
\frac{\omega^2}{c_0^2}\epsilon\mu, \label{eq:k}\EE where
$\epsilon$ and $\mu$ are respectively the frequency-dependent
permittivity and permeability relative to the vacuum. As shown
below, $k_x'$ equals the counter part in the vacuum: $k_x'=k_x$.
For the sake of convenience, hereafter, $k$ without a subscript
denotes the wavenumber in the $z$-direction, while with a
superscript `$\prime$' refers to the quantity inside the slab of
LHM.

Writing $ \epsilon = \epsilon_R + i\epsilon_I, \ \mbox{and} \ \mu
= \mu_R + i \mu_I, $ and taking them into Eq.~(\ref{eq:k}) we can
solve for $k'$ from \BA {k'}^2 + k_x^2
&=&k_0^2\left[(\epsilon_R\mu_R - \epsilon_I\mu_I) +
i(\epsilon_I\mu_R + \epsilon_R\mu_I)\right]\nonumber\\ &=& A +
iA\delta \label{eq:4} \EA where $$ k_0 \equiv
\frac{\omega}{c_0}^2, \ \ A \equiv k_0^2(\epsilon_R\mu_R -
\epsilon_I\mu_I), \ \ \delta \equiv \frac{\epsilon_I\mu_R +
\epsilon_R\mu_I}{\epsilon_R\mu_R - \epsilon_I\mu_I}.$$ The
solution to (\ref{eq:4}) can be written as \BE k' = k_R + ik_I.
\label{eq:5}\EE The physical requirement is that $k_I >0$, so not
to break the causality.

Taking Eq.~(\ref{eq:5}) into (\ref{eq:4}), we have $$ k_R^2 -
k_I^2 = A - k_x^2,\ \ \mbox{and} \ \ k_Rk_I = \frac{A\delta}{2}.
$$ From these equations we solve for $k_I$ and $k_R$ as \BA k_I
&=& \left( \frac{\sqrt{(A-k_x^2)^2 + A^2\delta^2} - (A-k_x^2)}{2}
\right)^{1/2},\label{eq:kI}\\ k_R &=&
\frac{A\delta}{2}\left(\frac{2}{\sqrt{(A-k_x^2)^2 + A^2\delta^2} -
(A-k_x^2)} \right)^{1/2}. \label{eq:kR}\EA The above procedure
determines the wavenumber (both the real and imaginary parts)
unambiguously. It also excludes the two possibilities (i) and
(iii) discussed in \cite{Garcia}. It is easy to verify that
without the the imaginary parts of the permittivity and
permeability the determination of the wavenumber is problematic;
therefore these imaginary pasts are essential to obtain the
correct form of the wavenumber.

The refractive index can be determined in the same procedure. We
just list the results. For $\epsilon_R\mu_R > \epsilon_I\mu_I$,
\BA n &=& \sqrt{\frac{\epsilon_R\mu_R -
\epsilon_I\mu_I}{2}}\nonumber\\ & & \times
\left(\frac{\delta}{\sqrt{\sqrt{1+\delta^2}-1}} +
i{\sqrt{\sqrt{1+\delta^2}-1}}\right).\nonumber \EA Obviously, when
$\epsilon_R = \mu_R = -1$ and $\epsilon_I, \mu_I << 1$, we have $n
\approx -1 + i\frac{|\delta|}{2} \approx = -1 +
i\left(\frac{\epsilon_I + \mu_I}{2}\right)$.

{\bf TE mode} We consider the TE mode, i.~e. {\bf E} is
perpendicular to the incidence plane, ${\bf E} = E\hat{y}$. The
electrical waves in the three regions can be expressed as $$
E_{\mbox{I}} = e^{ik_x x}(e^{ikz} + Re^{-ikz}),$$ $$ E_{\mbox{II}}
= e^{ik_x' x}(C e^{ik'z} + De^{-ik'z}),$$ $$ E_{\mbox{III}} =
e^{ik_x x} Te^{ikz}, $$ where $R$ and $T$ denote the reflection
and transmission coefficients respectively, and $C$ and $D$ are
coefficients for the waves inside the slab. These quantities are
determined by the boundary conditions. The corresponding magnetic
field ${\bf H}$ is determined by $ {\bf H} = \frac{1}{i\omega
\mu}\nabla \times {\bf E}. $ Therefore we have $$ H_{\mbox{I}, x}
= \frac{-k}{\omega \mu_1} e^{ik_x x}(e^{ikz} - R e^{-ikz}),$$ $$
H_{\mbox{II},x} = \frac{-k'}{\omega \mu_2} e^{ik_x' x} (Ce^{ik'z}
- D e^{-ik'z}),$$ $$ H_{\mbox{III},x} = \frac{-k}{\omega \mu_1}
e^{ik_x x} T e^{ikz},$$ where $\mu_1$ is the permeability outside
the slab, and $\mu_2$ is the permeability inside the slab.
Although the following derivations are good for any $\mu_1$ and
$\mu_2$, we restrict our attention to the case that the medium
outside the slab is the vacuum, and therefore $\mu_1 =1 $.

In the present case, the boundary conditions state that $k_x$, $E$
and $H_x$ are continuous across the boundaries at $z=0$ and $z=d$.
At $z=0$, the conditions imply $ k_x' = k_x,$ and \BA 1+R &=& C+D,
\nonumber\\ \frac{k}{\mu_1} (1-R) &=& \frac{k'}{\mu} (C-D).
\nonumber \EA These two equations can put into a matrix form \BE
\left(\begin{array}{c} 1 \\ R \end{array}\right) = \frac{1}{2}
\left(\begin{array}{cc} 1 + \frac{k'}{\mu k} & 1 - \frac{k'}{\mu
k} \\ 1 - \frac{k'}{\mu k} & 1 + \frac{k'}{\mu
k}\end{array}\right) \left(\begin{array}{c} C \\ D\end{array}
\right), \label{eq:mx1} \EE where $\mu = \frac{\mu_2}{\mu_1}$.
Here we note that the boundary conditions require $k_x=k_x'$, not
$k_x = -k_x'$. The mistake of $k_x = -k_x'$ made by many previous
authors has been pointed out in \cite{Valanju}.

Similarly at $z=d$, we obtain \BA Ce^{ik'd} + D e^{-ik'd} &=&
Te^{ikd}\nonumber\\ \frac{k'}{\mu_2}\left(Ce^{ik'd} -
De^{-ik'd}\right) &=& \frac{k}{\mu_1} T e^{ikd}.\nonumber \EA Or
putting\ into the matrix form, we have \BE \left(\begin{array}{c}
C \\ D
\end{array}\right) = \frac{T}{2} \left(
\begin{array}{c} (1+\frac{\mu k}{k'}) e^{i(k-k')d}\\ (1-\frac{\mu
k}{k'}) e^{i(k+k')d}\end{array}\right). \label{eq:mx2}\EE
Equations (\ref{eq:mx1}) and (\ref{eq:mx2}) lead to \BE
\left(\begin{array}{c} 1 \\ R \end{array} \right) =
\frac{Te^{ikd}}{4} \left(\begin{array}{cc} 1 + \frac{k'}{\mu k} &
1 - \frac{k'}{\mu k} \\ 1 - \frac{k'}{\mu k} & 1 + \frac{k'}{\mu
k}\end{array}\right) \left(
\begin{array}{c} (1+\frac{\mu k}{k'}) e^{-ik'd}\\ (1-\frac{\mu
k}{k'}) e^{ik'd}\end{array}\right)\label{eq:mx3}\EE This gives the
solution \BA T &=& \frac{4\mu k k' e^{-ikd}}{(\mu k + k')^2
e^{-ik'd} - (\mu k - k')^2 e^{ik'd}}, \label{eq:TET}\\ R &=&
\frac{2i[{k'}^2  - (\mu k)^2] \sin(k'd)}{(\mu k + k')^2e^{-ik'd} -
(\mu k - k')^2 e^{ik'd}}. \label{eq:TER}\EA Comparing to the
transmission coefficient $T_p$ obtained in \cite{Pendry}, the
present result differs by a fact of $\exp(ikd)$, i.~e. $ T = T_p
e^{ikd}.$

{\bf TM mode} For TM waves, the magnetic wave is perpendicular to
the incidence plane. And the electrical field is related to the
magnetic field as $ {\bf E} = \frac{-1}{i\omega \epsilon}
\nabla\times{\bf H}.$ Again we define $\epsilon =
\epsilon_2/\epsilon_1$, the corresponding quantities can be
obtained from that of the TE mode by the following substitution $$
{\bf E} \longleftrightarrow {\bf H},\ \ \ \mu \longleftrightarrow
\epsilon,\ \ \  \mu_{1,2} \longleftrightarrow -\epsilon_{1,2}.$$
Thus we have \BA T &=& \frac{4\epsilon k k' e^{-ikd}}{(\epsilon k
+ k')^2 e^{-ik'd} - (\epsilon k - k')^2 e^{ik'd}},
\label{eq:TMT}\\ R &=& \frac{2i[{k'}^2 - (\epsilon
k)^2]\sin(k'd)}{(\epsilon k + k')^2 e^{-ik'd} - (\epsilon k -
k')^2 e^{ik'd}}.\label{eq:TMR}\EA Again there is a fact of
$\exp(ikd)$ difference between the present result and that in
\cite{Pendry}.

The above results are derived rigorously and valid for any slab of
materials. However, when applied to LHM materials with
$\mu_R=\epsilon_R=-1$ and $\mu_I=\epsilon_I = 0$, there is an
obvious ambiguity. It is easy to verify that the transmitted
evanescent waves at the left interface of the slab go to infinity
as $d$ approaches $\infty$. And the portion inside the slab would
also go to infinity. This is against the results when we would
calculate the waves inside the slab by assuming a semi-infinite
slab in the beginning. Therefore $\mu_R=\epsilon_R=-1$ and
$\mu_I=\epsilon_I = 0$ cannot be satisfied simultaneously or these
parameters have to depend on the slab size. When $\mu_R,
\epsilon_R\neq -1$ or $\mu_I, \epsilon_I \neq 0$, the slab can
still amplify the evanescent waves. But as $d$ increases, the
transmission coefficient approaches a constant and the
amplification disappears, then the slab cannot restore the lost
information. It can be shown that roughly speaking, to get a
reasonable recovery of the evanescent waves $d$ should be
maximally around $1/k_0$, which severely limits the use of LHM
lenses.

Now we consider imaging by LHM lenses. Again we consider the two
dimensional model. Applying the above results to
Eq.~(\ref{eq:int}), the electrical wave arriving at the image site
will be \BE E(0,0,d) = \int_{-\infty}^\infty dk_x A(k_x,k)T
e^{i{k_x x+ 2ikd}}.\label{eq:int2}\EE The integration components
$\int_{\omega/c_0}^\infty$ and $\int_{-\infty}^{-\omega/c_0}$,
i.~e. the evanescent branches $k= i\sqrt{-(\omega/c_0)^2 + k_x^2},
\ \ \mbox{with}\ \ |k_x|
> k_0$ suffer a loss of $\exp(2ikd)$. The next question is whether
the transmission through the slab of LHM can fully overcome this
loss. If it could, then the slab would make a perfect lens. If not
fully, then why and to what extend it could. To answer these
questions, we have carried out a series of numerical simulations.

We will focus upon the evanescent waves, i.~e. $k_x > k_0$. A
recover rate can be defined as $ Q = e^{2ikd}T.$ Since $T$ is
normally a complex number, the actually waves arriving at the
image will also be subject to a phase shift caused by passing
through the lens. The phase shift is given by \BE \theta =
\tan^{-1}(T_I/T_R),\EE where $T_I$ and $T_R$ denote the imaginary
and real parts of $T$ respectively. A perfect lens would require
that \BE |Q| = 1, \ \ \ \theta = 0. \label{eq:re}\EE These can be
achieved when we take the unphysical limits in Eq.~(\ref{eq:TET})
$$
\begin{array}{c} \epsilon_R\\ \mu_R \end{array} \longrightarrow
-1, \ \ \ \mbox{and} \ \ \
 \begin{array}{c} \epsilon_I\\
\mu_I \end{array} \longrightarrow 0.$$ This agrees with the
conjecture of \cite{Pendry}. However, this situation violates the
causality. The absorption, however small, has to be present, or
$\mu_R$ and $\epsilon_R$ cannot be -1 at the same time. {\it Any
deviation from the identities in} Eq.~(\ref{eq:re}) {\it would
destruct perfect imaging.}

\begin{figure}[hbt]
\begin{center}
\epsfxsize=2.5in \epsffile{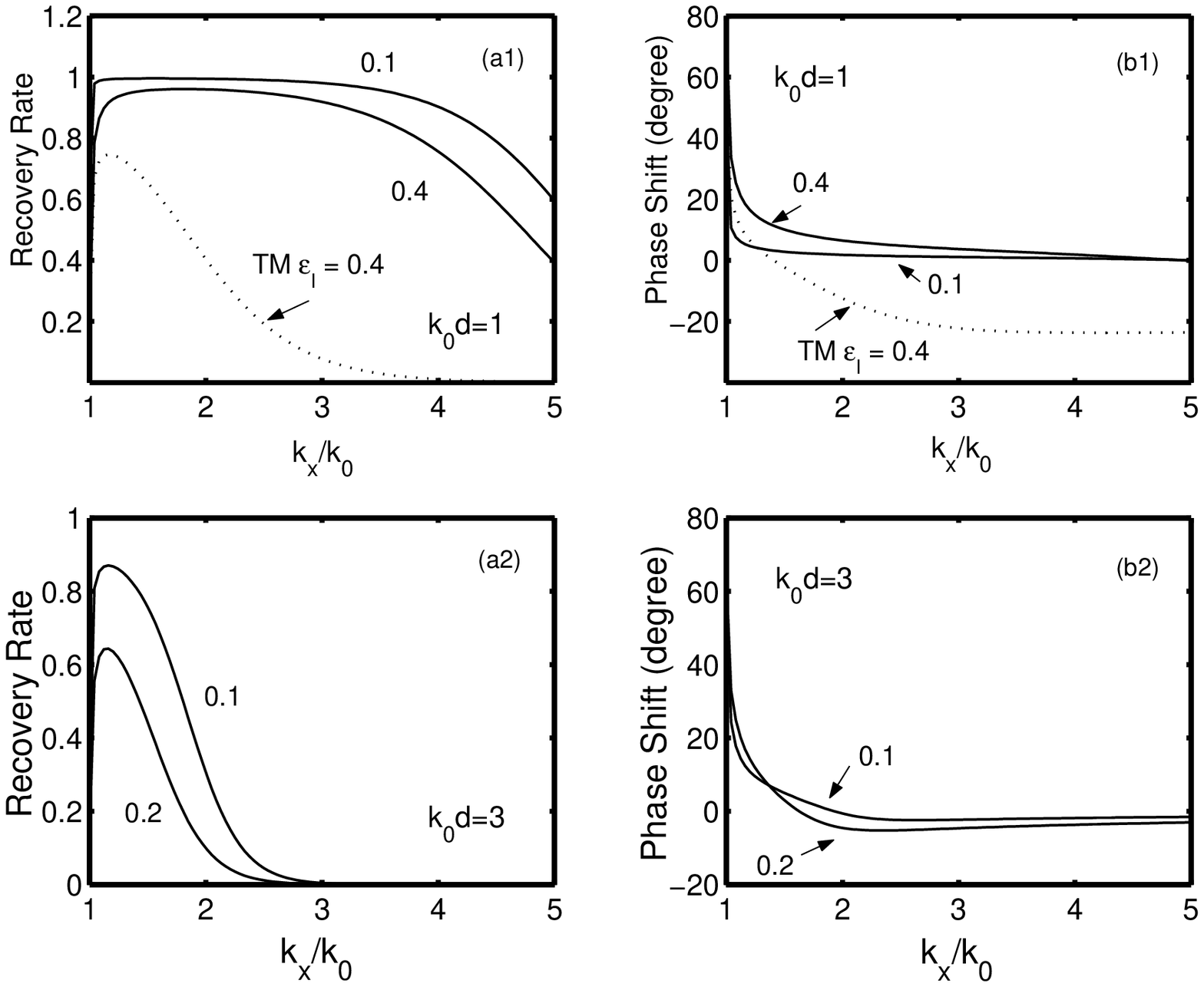} \caption{ \label{fig2}\small
The recover rate $Q$ and the phase shift of the evanescent waves
versus $k_x/k_0$. The imaginary part of the dielectric constant is
indicated in the figure. The results for the TE and TM modes are
represented by the solid and dotted line respectively. Here we
take $k_0d = 1$ for (a1) and (b1) and $k_0d=3$ for (a2) and (b2).}
\end{center}
\end{figure}

Fig.~\ref{fig2} shows the recovery rate and the phase shift as a
function of $k_x/k_0$. Here we use the following parameters:
$\epsilon_R = \mu_R = -1$, $\mu_I=0.001$. The values for
$\epsilon_I$ are taken as 0.4 and 0.1 for $k_0d=1$ and 0.1 and 0.2
for $k_0d=3$, corresponding roughly to the refraction index $n =
-1 + 0.2i$, $n = -1 + 0.05i$ and $n=-1+0.1i$ respectively. The
value $\epsilon_I = 0.4$ and $0.2$ were from \cite{Pendry} and
\cite{Garcia} separately.

What have been shown are as follows. (1) The lens by LHM does
amplify the evanescent waves, in disagreement with \cite{Garcia}
but in agreement with \cite{Pendry}. However, (2) even with a
small amount of absorption, represented by the imaginary parts of
permittivity and permeability, the information around $k_0$ still
suffers a significant distortion after passing the amplifying
lens. The frequency range within which the lost information can be
recovered is not wide even for small $\epsilon_I = 0.1$. The
absorption plays a crucial role in reducing the ability of the
lens to make an image. (3) The range of good recovery increases
with vanishing $\epsilon_I$. (4) Although the evanescent waves are
recovered to a certain degree, the question of whether such a
recovery would improve or worsen the imaging resolution is still
unclear at this stage. (5) The recovery rate is worse for the TM
mode compared to the TE mode with the same value of $\epsilon_I$.
Note that the TM mode with $\epsilon_I =0.4$ was the case that has
been considered in \cite{Pendry}. (6) Given the poor recovery rate
and the significant phase shift, whether there is any benefit for
using the LHM lens to improve the resolution of a traditional lens
is uncertain. (7) When the slab width $d$ increases, even just
slightly bigger than a few folds of the wavelength, the recovery
rate will drop drastically as shown in (a2). This will severely
limit the use of LHM lenses for imaging.

\input{epsf}

\begin{figure}[hbt]
\begin{center}
\epsfxsize=2.5in \epsffile{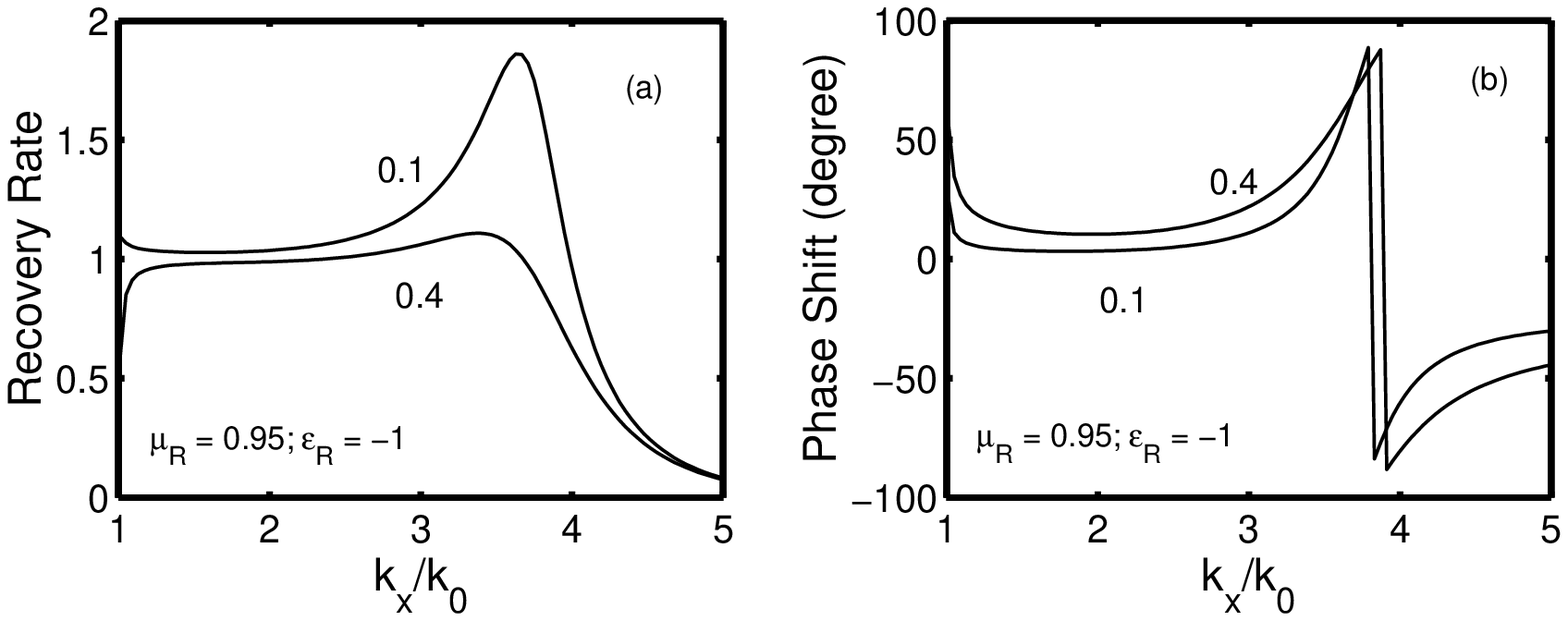} \caption{ \label{fig3}\small
The TE mode. The recover rate and the phase shift of the
evanescent waves versus $k_x/k_0$ for $\mu_R=-0.95$ and
$\epsilon_R = -1$. The imaginary part of $\epsilon$ is indicated
in the figure. In the simulation $k_0d = 1$.}
\end{center}
\end{figure}

We have also investigated the sensitivity of the recovery rate to
the real part of permittivity and permeability. We found that it
is less sensitive to permittivity than to permeability for TE
waves; in the case of TM waves, the situation reverses. In
Fig.~\ref{fig3}, we plot the recovery rate and the phase shift for
the TE mode as a function of $k_x/k_0$ for $\epsilon_R = -1$ and
$\mu_R = -0.95$. Again two values were chosen for $\epsilon_I$. It
is shown that interestingly, the quality of recovery does not
always increase with vanishing $\epsilon_I$. At certain frequency
region, the evanescent waves are over amplified. And such an
amplification increases with decreasing absorption. At the extreme
case, $\mu_I=\epsilon_I = 0$, the amplification at about $k_x/k_0
= 3.75$ can be as large as 100. Comparing the results in
Fig.~\ref{fig2} and Fig.~\ref{fig3}, we may conclude that the
variation in the real part of the permittivity and permeability
are more crucial to imaging. Since these two parameters are
frequency sensitive, a care must be taken when making images using
non-monochromatic waves.

In summary, we have evaluated the quality of lenses made by left
handed materials using the standard transfer matrix method. It is
shown that although the evanescent waves can be amplified by LHM
made lenses, the necessary presence of absorption makes the lenses
not only much less perfect, but practically useless. Some
ambiguities in previous investigations have also been mentioned.

\input{epsf}

\end{document}